\documentclass[aps,prd,amsmath,preprint]{revtex4}

\usepackage{enumerate}
\usepackage{amsmath,amssymb}
\usepackage{bm}
\usepackage{amscd}            
\usepackage{graphicx}
\usepackage{hhline,multirow}  
\usepackage{dcolumn}          
\usepackage{url}              
\usepackage{color}
\newcommand{\vecp}{\bm p}

\preprint{JLAB-THY-14-1874}

\begin{document}
\title{Nuclear effects in the proton--deuteron Drell-Yan process}

\author{P.~J.~Ehlers$^{1,2}$,
        A.~Accardi$^{2,3}$,
	L.~T.~Brady$^{2,3}$%
	\footnote{Current address:
	University of California, Santa Barbara, California 93106, USA},
	W.~Melnitchouk$^2$}
\affiliation{
$^1$\mbox{University of Minnesota -- Morris, Morris, Minnesota 56267, USA}
$^2$\mbox{Jefferson Lab, Newport News, Virginia 23606, USA}
$^3$\mbox{Hampton University, Hampton, Virginia 23668, USA} \\
}

\date{\today}

\begin{abstract}
We compute the nuclear corrections to the proton--deuteron Drell-Yan
cross section for inclusive dilepton production, which, when combined
with the proton--proton cross section, is used to determine the flavor
asymmetry in the proton sea, $\bar d \neq \bar u$.
In addition to nuclear smearing corrections that are known to be
important at large values of the nucleon's parton momentum fraction
$x_N$, we also consider dynamical off-shell nucleon corrections
associated with the modifications of the bound nucleon structure
inside the deuteron, which we find to be significant at intermediate
and large $x_N$ values.
We also provide estimates of the nuclear corrections at kinematics
corresponding to existing and planned Drell-Yan experiments at
Fermilab and J-PARC which aim to determine the $\bar d/\bar u$ ratio
for $x \lesssim 0.6$.
\end{abstract}

\maketitle

\section{Introduction}

The discovery of the flavor asymmetry in the light quark sea
in the proton, $\bar d \neq \bar u$, has been one of the most
important findings in hadronic physics from the past two decades
\cite{NMC91, NMC94, HERMES, NA51, E866_98, E866},
stimulating considerable discussion about the nature and origin
of the nucleon's nonperturbative structure (see {\it e.g.}
Refs.~\cite{Kumano98, Speth98, Garvey01, Peng14} for reviews).
In particular, the Drell-Yan reaction, involving dilepton pair
production in inclusive hadron--hadron scattering, has provided
the most direct constraints on the $x$ dependence of the
$\bar d/\bar u$ ratio \cite{NA51, E866_98, E866}.
Here, at the partonic level, a quark from the hadron beam
annihilates with an antiquark from the target hadron
(or vice versa), producing a high energy virtual photon which
subsequently decays to a pair of oppositely charged leptons,
$q \bar q \to \gamma^* \to \ell^+ \ell^-$ \cite{DY70}.
By selecting specific values of the momentum fractions of the
partons in the beam and target hadrons, one can construct
ratios of cross sections with sensitivity to particular
combinations of parton distribution functions (PDFs).
In contrast to inclusive deep-inelastic lepton--nucleon scattering,
which measures charge-even combinations of PDFs, $q+\bar q$,
the Drell-Yan reaction has the advantage of allowing effects
in the antiquark distributions to be cleanly isolated from those
in the quark PDFs.

More specifically, the inclusive proton--free nucleon scattering
cross section for the production of a lepton pair with invariant
mass squared $Q^2 \gg M^2$, where $M$ is the mass of the nucleon,
is given (at leading order in the strong coupling) by
\begin{equation}
\label{eq:pnxs}
\sigma^{pN}(x_p,x_N)\ \equiv\ \frac{d\sigma^{pN}}{dx_p dx_N}
= \frac{4\pi\alpha^2}{9 Q^2}
  \sum_q e^2_q 
  \Big[ q(x_p)\bar{q}(x_N) + \bar{q}(x_p)q(x_N) \Big],
\end{equation}
where $\alpha$ is the fine-structure constant.
In Eq.~(\ref{eq:pnxs}), $q(x_p)$ and $\bar q(x_N)$ are the quark
and antiquark PDFs in the proton and target nucleon, evaluated
at parton light-cone momentum fractions $x_p$ and $x_N$ of the
proton and nucleon, respectively, $e_q$ is the electric charge,
and the sum is taken over all flavors $q$.
(Here and throughout this paper, for ease of notation we omit
the explicit $Q^2$ dependence in the arguments of PDFs and
cross sections.)
For the case of proton scattering from the deuteron, taking the
ratio of $pd$ to $pp$ cross sections, and assuming the deuteron
to be composed of a free proton and neutron, one can isolate the
ratio of $\bar d$ to $\bar u$ distributions for large values of
$x_p \gg x_N$ \cite{Ellis91},
\begin{equation}
\frac{\sigma^{pd}}{\sigma^{pp}}
\approx 1 + {\bar{d}(x_N) \over \bar{u}(x_N)}
	\hspace{80 pt} [x_p \gg x_N].
\label{eq:approxratio}
\end{equation}
The increase of the ratio $\sigma^{pd}/2\sigma^{pp}$ above unity
observed at intermediate $x$ values \cite{NA51, E866_98, E866} has
then been related directly to an excess of $\bar d$ over $\bar u$
in the proton.

Previous analyses of lepton-pair production in $pd$ scattering have
typically assumed that effects associated with the nuclear structure
of the deuteron are negligible at the energies where the existing
experiments \cite{NA51, E866_98, E866} have been carried out.
On the other hand, there has been a growing awareness of the need to
account for nuclear corrections in precision determinations of PDFs,
particularly at large values of $x$ \cite{AKL04, KP06, Kahn09, AQV},
where there is greatest sensitivity to the short-range structure of
the nucleon--nucleon interaction.
In deep-inelastic lepton scattering from the deuteron, for instance,
nuclear smearing and nucleon off-shell effects have been included
in a number of global PDF analyses \cite{ABKM09, CJ10, CJ11, CJ12,
Brady12, JMO13}, which have found significant effects on the $d$-quark
distribution in particular at high $x$ values ($x \gtrsim 0.5$).
While the sea quark distributions in the proton do not extend to as
large values of $x$ as the valence quark distributions, the smearing
effects become prominent at correspondingly smaller $x$ values where
the PDFs are falling most rapidly \cite{MST98}.

Recently Kamano and Lee \cite{KL12} considered nuclear corrections
to the $pd$ Drell-Yan cross sections in a center of mass frame in
which the projectile and target move with large longitudinal momentum.
Boosting the deuteron wave function from the rest frame under the
assumption that particle number is conserved, they found small
corrections to the existing data from the Fermilab E866 experiment
\cite{E866_98, E866}, but potentially larger effects at $x \gtrsim 0.5$
for the new E906 (``SeaQuest'') experiment at lower energy \cite{E906}.
Contributions from pion exchange between the proton and neutron in
the deuteron were also found to be important at $x \gtrsim 0.4$
\cite{KL12}, which may appear surprising given that the small mass
of the pion $m_\pi$ is generally expected to restrict such effects
to the region of $x \lesssim m_\pi/M \approx 0.15$.
Earlier calculations of pion (and other meson) exchange effects in
deep-inelastic lepton--deuteron scattering \cite{Kaptari91, MT93,
Nikolaev97} showed a few percent overall enhancement (``antishadowing'')
in the deuteron to nucleon structure function ratio at $x \sim 0.1$.
An indirect feedback effect on quark distributions at large $x$
could result from the conservation of valence quark number,
although one might expect the major impact of the pion cloud
to be on sea quark distributions.

In addition to Fermi motion and meson exchange effects, other
corrections are also known to contribute to high energy nuclear
cross sections, such as those associated with nuclear medium
modification of the partonic structure of bound nucleons (nucleon
off-shell corrections) \cite{GL92, Ciof92, MST94, MSTplb, KPW94},
and final state interactions between the spectator nucleon in
the deuteron and the hadronic debris from the proton--nucleon
collision \cite{Cosyn13}.
While both of these corrections are difficult to constrain
theoretically, their uncertainties are important to estimate
for determining the overall errors on the PDF distributions,
especially at large values of $x$.

In the present analysis we revisit the calculation of the
proton--deuteron Drell-Yan cross section in Sec.~\ref{sec:pd},
paying particular attention to corrections associated with
Fermi smearing and nucleon off-shell effects, which are
expected to persist even at high energies.
These corrections have received attention recently in several
analyses \cite{ABKM09, CJ10, CJ11, CJ12} of deep-inelastic
scattering from the deuteron, where their impact on PDFs and
their uncertainties have been studied systematically in the
context of global QCD fits using the ``weak binding approximation''
\cite{AKL04, KP06, Kahn09}.
However, to date the global PDF analyses have not systematically
included nuclear corrections to the $pd$ Drell-Yan data, and
it is important for a consistent determination of PDFs to
consistently incorporate these in all data sets analyzed which
involve deuterium nuclei.

In Sec.~\ref{sec:pN} we evaluate the $pN$ cross section in terms of
parton distributions in the bound nucleon.  To take into account the
possible modification of the nucleon structure due to interactions
with the nuclear environment, we consider a relativistic spectator
quark model in which the bound nucleon PDF is related to the change of
the confinement radius of the nucleon in the deuteron.  The approach
is similar to the model developed in Refs.~\cite{KP06, CJ12}, but in
addition to valence quarks the model also includes the effects on sea
quarks at small $x$.
The combined effects of the nuclear smearing and off-shell corrections
are illustrated in Sec.~\ref{sec:results}, where we discuss their
impact on existing and planned Drell-Yan experiments at Fermilab 
\cite{E866_98, E866, E906} and J-PARC \cite{J-PARC, Kumano10}.
Finally, in Sec.~\ref{sec:conc} we summarize our findings, and discuss         
their implications for future analyses of Drell-Yan cross sections
and their constraints on parton distributions.

\section{Drell-Yan process in proton--deuteron scattering}
\label{sec:pd}

In this section we present the derivation of the inclusive lepton
pair production cross section for proton--deuteron scattering.
After defining the relevant light-cone kinematics in the collinear
frame, we describe how the $pd$ cross section can be related to the
corresponding nucleon-level cross section for proton scattering
from bound nucleons in the deuteron.

\subsection{Kinematics}
\label{ssec:kinematics}

We begin by defining the four-momenta of the virtual photon,
beam proton, and target deuteron by $q^\mu$, $k^\mu$, and $P_d^\mu$,
respectively.  To simplify the notation, we also introduce a rescaled
deuteron momentum, $p_d^\mu \equiv (M / M_d) P_d^\mu$, where $M_d$ is
the deuteron mass, so that $p_d^2 = M^2$.
In a frame of reference where the proton and deuteron are collinear
(``$pd$ frame'') the momenta can be decomposed in terms of light-cone
unit vectors $\bar n^\mu$ and $n^\mu$,
\begin{subequations}
\label{eq:lcdecomp}
\begin{align}
  q^\mu   & = x_p\, k^+\, \bar n^\mu
	    + x_d\, p_d^-\, n^\mu
	    + q_\perp^\mu,			\label{eq:qmu} \\
  k^\mu   & = k^+\, \bar n^\mu
	    + \frac{M^2}{2 k^+} n^\mu,		\label{eq:ppmu} \\
  p_d^\mu & = \frac{M^2}{2p_d^-}\, \bar n^\mu
	    + p_d^-\, n^\mu,			\label{eq:pdmu}
\end{align}
\end{subequations}
where $\bar n^2=n^2=0$ and $\bar n \cdot n = 1$, and
the ``plus'' and ``minus'' light-cone components of any
four-vector $a$ are defined as $a^\pm = (a_0 \pm a_3)/\sqrt{2}$.
The four-vector $q_\perp^\mu$ denotes the transverse momentum of
the photon, with $q_\perp \cdot n = q_\perp \cdot \bar n = 0$,
and ${\bm q}_\perp^2 = -q_\perp^\mu {q_{\perp}}_\mu$ the square
of the corresponding transverse three-momenta.
The four-momentum of the nucleon in the deuteron is denoted
by $p^\mu$ and can be similarly expanded as
\begin{align}
  p^\mu & = \frac{p^2+{\bm p}_\perp^2}{2 z p_d^-} \bar n^\mu
	  + z p_d^-\, n^\mu
	  + p_\perp^\mu,			\label{eq:pNmu}
\end{align}
where the transverse momentum four-vector $p_\perp^\mu$
of the bound nucleon is defined such that
${\bm p}_\perp^2 = -p_\perp^\mu p_{\perp \mu}$.
The variables $x_p$ and $x_d$ in Eq.~(\ref{eq:qmu})
are given by
\begin{align}
  x_p = \frac{q^+}{k^+}, \qquad
  x_d = \frac{q^-}{p_d^-},
\label{eq:Nachtmann}
\end{align}
and play the role of Nachtmann scaling variables for the
Drell-Yan process, while the variable $z$ in Eq.~(\ref{eq:pNmu})
represents the light-cone momentum fraction carried by the nucleon
in the deuteron,
\begin{align}
  z = \frac{p^-}{p_d^-}.
\label{eq:z}
\end{align}
Note that in the collinear frame, where the beam and target move
in opposite directions, the proton and deuteron variables involve
``plus'' and ``minus'' components, respectively.

The experimentally measured (external) variables characterizing
the process are the rescaled center of mass energy squared of the
collisions, $s = (k + p_d)^2$, the dilepton invariant mass squared
$Q^2 = (\ell + \bar\ell)^2$, where $\ell$ and $\bar\ell$
are the four-momenta of the produced lepton and antilepton.
For convenience we also define the variable
$\tilde s = 2 k^+ p_d^-$, in terms of which the center
of mass energy squared can be written as
$s = \tilde s + 2 M^2 + M^4/\tilde s$, so that in the high
energy limit, $\tilde s \gg M^2$, one has $s \to \tilde s$.
From Eq.~(\ref{eq:qmu}) the photon virtuality can also be
related to $\tilde s$ as
$Q^2 = x_p x_d\, \tilde s - {\bm q}_\perp^2$.
In addition, one can define the dilepton rapidity $y^*$ in the
proton--deuteron center of mass frame (or ``$pd$ frame''),
in which $k^+ = p_d^-$,
\begin{align}
  y^* & = \frac12 \log \frac{q^+}{q^-}\bigg|_{k^+ = p_d^-}.
\end{align}
Note that, in contrast to the Lorentz invariants $s$ and $Q^2$,
the rapidity generally depends on the frame of reference.
The external variables are related to the Nachtmann light-cone
momentum fractions by
\begin{align}
  x_p = \frac{Q_\perp}{\sqrt{\tilde s}} \exp(y^*),
  \qquad
  x_d = \frac{Q_\perp}{\sqrt{\tilde s}} \exp(-y^*),
\end{align}
where $Q_\perp^2 = Q^2 + {\bm q}_\perp^2$ is the transverse mass
of the dilepton, and inverting the relation between $s$ and
$\tilde s$, one has
\begin{align}
  \tilde s
  & = \frac{s}{2}
      \left[ 1 - {2M^2 \over s} + \sqrt{1 - {4M^2 \over s}}
      \right].
\end{align}

In the nuclear impulse approximation one assumes that the proton beam
scatters incoherently from the individual proton or neutron in the
deuteron.  In relating the $pd$ Drell-Yan cross section to the
underlying proton--bound nucleon cross section, it will be convenient
to introduce the (internal) nucleon level analog of the Nachtmann
scaling variable,
\begin{align}
  x_N & = \frac{q^-}{p^-} = \frac{x_d}{z},
\label{eq:xN}
\end{align}
the $pN$ center of mass energy squared,
\begin{align}
  s_N & = (k + p)^2 
	= z \tilde s
	  \left[ 1 + \frac{M^2 + p^2 + {\bm p}_\perp^2}{z \tilde s}
		   + \frac{M^2 (p^2 + {\bm p}_\perp^2)}{(z\tilde s)^2}
	  \right],
\label{eq:sN}
\end{align}
and the nucleon level rapidity,
\begin{align}
  y_N^*&= \frac12 \log\frac{q^+}{q^-}\bigg|_{k^+ = p^-}
	= y^* + \log \sqrt{z},
\label{eq:yN}
\end{align}
where each of the variables has also been related to the external
variables defined above.
Note that the rapidity $y_N^*$ here does not coincide with the
rapidity of a free proton--bound nucleon collision, since the
center of mass is slightly shifted by the transverse component
of the nucleon momentum \cite{Mul01}.
While the kinematical relations in Eqs.~(\ref{eq:xN})--(\ref{eq:yN})
are exact, in practice the energies relevant for current and future
Drell-Yan experiments are relatively large, with $s \gg M^2$.
In the high energy limit, $s \to \infty$, one can therefore
usually neglect hadron mass corrections and the transverse motion
of the nucleon, in which case the proton and deuteron momentum
fractions simplify to
\begin{align}
  x_p & \approx \frac{Q_\perp}{\sqrt{s}} \exp(y^*),	\qquad
  x_d \approx \frac{Q_\perp}{\sqrt{s}} \exp(-y^*),
\end{align}
while the $pN$ center of mass energy squared becomes
$s_N \approx z s$.
For the inclusive cross sections that will be considered here,
with ${\bm q}_\perp^2$ integrated over, one can also assume
$Q^2 \gg {\bm q}_\perp^2$, so that $Q_\perp \approx Q$.

\subsection{Relation between deuteron and nucleon cross sections}
\label{ssec:crosssec}

The differential cross section for Drell-Yan lepton pair production
in inclusive proton--deuteron scattering is defined as \cite{Mul01}
\begin{align}
  { d\sigma^{pd} \over d^4q \, d\Omega }
    = \frac{1}{4\sqrt{(k \cdot p_d)^2 - M^4}}
    \frac{\alpha^2}{Q^4}
    L^{\mu\nu}(\ell,\bar \ell)\, W^{pd}_{\mu\nu}(k,p_d,q),
\label{eq:sigmapd-def}
\end{align}
where $\Omega$ is the solid angle spanned by the lepton pair.
The lepton tensor $L^{\mu\nu}$ is given by
\begin{equation}
\label{eq:lepton}
  L^{\mu\nu}(\ell,\bar \ell)
    = 2 \ell^\mu \bar \ell^\nu
    + 2 \ell^\nu \bar \ell^\mu
    - g^{\mu\nu} (\ell \cdot \bar{\ell} + m_\ell^2),
\end{equation}
where $m_\ell$ is the lepton mass, which in our case is negligible.
The proton--deuteron hadronic tensor (rescaled per-nucleon) is
defined in terms of the matrix element of the commutator of the
currents $J_\mu$ evaluated at the space-time points 0 and $\zeta$,
\begin{align}
  W^{pd}_{\mu\nu}(k,p_d,q) 
    = \frac{M}{M_d} \int\!\frac{d^4\zeta}{(2\pi)^4} e^{iq\cdot\zeta}
    \langle k,p_d | [J_\mu(0),J_\nu(\zeta)] | k,p_d \rangle,
\end{align}
with the arguments defined in Sec.~\ref{ssec:kinematics}.

In the weak binding approximation, as is relevant for a weakly bound
nucleus such as deuterium, the nucleon propagator in the nuclear
medium can be expanded up to order ${\bm p}^2/M^2$ in the bound
nucleon momentum \cite{KPW94, KMPW95, KMd}.  This then allows the
deuteron tensor to be factorized into a nucleon level tensor
${\widetilde W}^{pN}_{\mu\nu}$ and a deuteron spectral function
$\rho_d$ which describes the momentum distribution of the nucleons
in the deuteron,
\begin{align}
  W^{pd}_{\mu\nu}(k,p_d,q) 
    = \sum_N \int\!\frac{d^4p}{(2\pi)^4}\,
      \rho_d(p)\, {\widetilde W}^{pN}_{\mu\nu}(k,p,q),
\end{align}
where the sum is taken over the proton and neutron, $N=p+n$, and
we assume charge symmetric nucleon distributions in the deuteron.
A similar factorization can be obtained if one neglects antiparticle
degrees of freedom, or treats the nucleons effectively as scalars
\cite{AQV}.
In the impulse approximation, where the scattering takes place
incoherently from individual nucleons in the nucleus, with the
noninteracting ``spectator'' nucleon on its mass shell, the
spectral function can be written in terms of the deuteron's
rest frame wave function $\psi_d$ (which is a function of the
nucleon's three-momentum only),
\begin{align}
  \rho_d({\bm p})
    = {\cal N}\, (2\pi)^4 \left| \psi_d(\vecp) \right|^2
      \delta(p_0 + E_s - M_d),
\label{eq:rho_d}
\end{align}
where ${\cal N}$ is a normalization factor,
$E_s = \sqrt{ M^2 + {\bm p}^2 }$ is the spectator nucleon energy,
and
\begin{equation}
p_0\, =\, M_d - E_s\,
\approx\, M + \varepsilon_d - {{\bm p}^2 \over 2M}
\end{equation}
is the energy of the interacting nucleon, with 
$\varepsilon_d = -2.2$~MeV the deuteron binding energy.
In fact, the wave function depends on the magnitude of the
nucleon's three-momentum, $|\bm p|$, and is normalized such
that $\int d^3{\bm p} \left| \psi_d(\vecp) \right|^2 = 1$.

The specification of the deuteron rest frame in computing the
spectral function in Eq.~(\ref{eq:rho_d}) breaks the relativistic
covariance of the formalism, although typically Drell-Yan
experiments are performed with the deuteron target at rest
\cite{E866_98, E866, E906, J-PARC}.
Furthermore, the commonly used deuteron wave functions
\cite{Paris, AV18, CDBonn} are computed in the nonrelativistic
approximation, and care must be taken to ensure that the correct
normalization is preserved when reducing the full deuteron tensor,
defined in terms of relativistic nucleon fields, to one expressed
in terms of nonrelativistic wave functions.
(Relativistic extensions of deuteron wave functions, which
incorporate lower components of nucleon spinors, have also been
used recently in high precision fits to $NN$ scattering data
\cite{WJC}.)
Although the form of the normalization factor ${\cal N}$ is not unique
\cite{Ciof92, FS78}, the choice ${\cal N} = M/p_0$ ensures conservation
of the (Lorentz invariant) baryon number \cite{KPW94, KMPW95},
in contrast to the conservation of the (Lorentz non-invariant)
particle number, discussed in Ref.~\cite{KL12}.  With this choice,
the differential $pd$ cross section in Eq.~(\ref{eq:sigmapd-def})
can be written in the deuteron rest frame as 
\begin{equation}
\label{eq:pdxsgen}
  \frac{d\sigma^{pd}}{dx_p dx_d d^2{\bm q}_\perp d\Omega}
     = \sum_N \int\!d^3\vecp
       \frac{M}{p_0} |\psi_d(\vecp)|^2 
       \frac{\tilde s}{8M|\vecp|}
       \frac{\alpha^2}{Q^4}\,
       L^{\mu\nu}\, {\widetilde W}^{pN}_{\mu\nu}(k,p,q),
\end{equation}
where we have used the relation
$d^4q = k^+ p_d^-\, dx_p\, dx_d\, d^2{\bm q}_\perp$.

In analogy with Eq.~(\ref{eq:sigmapd-def}), the product of the
lepton tensor with the $pN$ hadronic tensor in Eq.~(\ref{eq:pdxsgen})
is related to the proton--off-shell nucleon differential scattering
cross section according to
\begin{equation}
\label{eq:pNxsgen}
  \frac{d{\widetilde\sigma}^{pN}}{dx_p dx_N d^2{\bm q}_\perp d\Omega}
    = \frac{z \tilde s}{8 \sqrt{(k \cdot p)^2 - M^2 p^2}}
      \frac{\alpha^2}{Q^4}
      L^{\mu\nu}\, {\widetilde W}^{pN}_{\mu,\nu}(k,p,q),
\end{equation}
where $dx_N = dx_d/z$.
Integrating over the transverse photon momentum ${\bm q}_\perp$
and $d\Omega$, the proton--deuteron cross section
$\sigma^{pd}(x_p,x_d) \equiv d\sigma^{pd}/dx_p dx_d$ can therefore
be written in terms of the $pN$ cross section as
\begin{equation}
\label{eq:pdwithpN}
  \sigma^{pd}(x_p,x_d) 
    = \sum_N \int\!d^3\vecp\, |\psi_d(\vecp)|^2\,
      \frac{\sqrt{(k \cdot p)^2 - M^2 p^2}}{z p_0 |\vecp|}\,
      {\widetilde\sigma}^{pN}\Big(x_p,\frac{x_d}{z},p^2\Big),
\end{equation}
where ${\widetilde\sigma}^{pN}
	\equiv d{\widetilde\sigma}^{pN}/dx_p dx_N$.
Note that the proton--nucleon cross section here has an explicit
dependence on $p^2$, which reflects the possible modification of
the nucleon structure due to interactions with other nucleons in
the nucleus.  This will require a generalization of the expression
for the on-shell cross section in Eq.~(\ref{eq:pnxs}) to account
for the dependence of the bound nucleon PDFs on $p^2$, which in
general does not vanish even in the high energy limit.
For the case of the deuteron, where the spectator nucleon is
on-mass-shell, the virtuality of the interacting nucleon $p^2$
can be related to ${\bm p}_\perp^2$ and $z$ by
$p^2 = -{\bm p}_\perp^2 M_d/(M_d-M z) + p^2_{\rm max}$, with
$p^2_{\rm max} = z M (M_d^2 - M^2 - z M M_d)/(M_d - M z)$.
The off-shell generalization of the $pN$ cross section and the
parton distributions in the bound nucleon will be discussed in
Sec.~\ref{ssec:offshell}.

\subsection{Convolution and nuclear smearing function}
\label{sec:smearing}

In practical applications of the cross section relation in
Eq.~(\ref{eq:pdwithpN}) it is convenient to express the
three-dimensional integration over ${\bm p}$ in terms of
integrations over the light-cone fraction $z$ and the transverse
momentum $p_\perp^2$, as defined in Sec.~\ref{ssec:kinematics},
\begin{equation}
\label{eq:jacobian}
  \int d^3\vecp
    = \int dz\, d\vecp^{\text{ }2}_{\perp}
      \frac{\pi M E_s}{M_d - M z},
\end{equation}
where the integration over the azimuthal angle has been performed.
The on-shell spectator condition restricts the nucleon momentum
to be ${\bm p}^2 = {\bm p}_\perp^2 + p_z^2$, where the longitudinal
momentum is given by
$p_z = \big[{\bm p}_\perp^2 + M^2 - (M_d-M z)^2\big]/2(M_d-M z)$.
Using these relations, the $pd$ cross section can then be written as
\begin{eqnarray}
\label{eq:pdxsfull}
  \sigma^{pd}(x_p,x_d)
    = \sum_N \int\!\frac{dz}{z} \, d\vecp^{\text{ }2}_{\perp}\,
      f(z,\vecp^2_{\perp})\,
      {\widetilde\sigma}^{pN}\Big(x_p,{x_d\over z},p^2\Big),
\end{eqnarray}
where 
\begin{align}
\label{eq:tildef}
  f(z,\vecp^2_{\perp})
    & = \frac{\pi M E_s}{M_d - M z}\,
    \frac{\sqrt{(k_0 p_0 + |k_z| p_z)^2 - M^2 p^2}}
	       {|k_z| p_0}\,
    |\psi_d(\vecp)|^2
\end{align}
is the $z$- and ${\bm p}_\perp$-dependent light-cone momentum
distribution of nucleons in the deuteron (or unintegrated smearing
function).
In the high energy limit, as is considered in the applications here,
where $k_0^2 \gg p^2 + {\bm p}_\perp^2
             \lesssim {\cal O}(1~{\rm GeV}^2)$,
one can approximate the function $f(z,\vecp^2_{\perp})$ by
\begin{align}
  f(z,\vecp^2_{\perp})
     \approx  
     \frac{\pi M E_s}{M_d - M z}\,
     \Big( 1 + \frac{p_z}{p_0} \Big)\, |\psi_d(\vecp)|^2.
\end{align}
This result coincides with the smearing function for deep-inelastic
scattering computed in the Bjorken limit \cite{KP06, AQV}, and is
automatically normalized to unity.
A simplified convolution in terms of one-dimensional smearing
functions can be obtained if the off-shell nucleon cross section
is expanded around its on-shell limit, $p^2 = M^2$,
\begin{equation}
  {\widetilde\sigma}^{pN}(x_p,x_N,p^2)
  \approx
    \sigma^{pN}(x_p,x_N)
    \left[ 1\, + \, \frac{(p^2-M^2)}{M^2}\,\, \delta\sigma^{pN}(x_p,x_N)
    \right],
\label{eq:sig_off}
\end{equation}
where $\sigma^{pN}(x_p,x_N)
       \equiv {\widetilde\sigma}^{pN}(x_p,x_N,M^2)$
is the on-shell proton--nucleon cross section, and
\begin{align}
  \delta\sigma^{pN}(x_p,x_N)
    = \left.
      \frac{\partial\log{\widetilde\sigma}^{pN}}{\partial \log p^2}
      \right|_{p^2=M^2}
\end{align}
is the lowest order off-shell correction.  Higher order terms
in this expansion are suppressed by additional powers of
$(p^2-M^2)/M^2 \approx 2 (\varepsilon_d - {\bm p}^2/M)/M$.
The expansion (\ref{eq:sig_off}) then enables the ${\bm p}_\perp$
dependence of the integrand to be factorized into a
${\bm p}_\perp$-integrated (on-shell) smearing function $f(z)$
and an off-shell smearing function $f^{(\rm off)}(z)$,
\begin{equation}
\label{eq:pdxs_smeared}
  \sigma^{pd}(x_p,x_d)
  = \sum_N \int_{x_d}^1 \frac{dz}{z}
    \Big[ f(z) + f^{(\text{off})}(z)\,
	  \delta\sigma^{pN}\Big(x_p,{x_d \over z}\Big)
    \Big]\,
    \sigma^{pN}\Big(x_p,{x_d \over z}\Big),
\end{equation}
where
\begin{align}
\label{eq:fz}
  f(z) & = \int\!d\vecp^2_{\perp}\, f(z,\vecp^2_{\perp}),
\end{align}
and
\begin{align}
\label{eq:foffz}
  f^{(\text{off})}(z) & = \int\!d\vecp^2_{\perp}\,
    \frac{p^2-M^2}{M^2} f(z,\vecp^2_{\perp}).
\end{align}
The result (\ref{eq:pdxs_smeared}) is analogous to the generalized
convolution expressions for deep-inelastic nuclear structure functions
\cite{KP06} in the high energy limit, corresponding in particular to
the $F_1^d$ structure function rather than $F_2^d$ (or $xF_1^d$).

\begin{figure}[t]
\includegraphics[width=7.8cm]{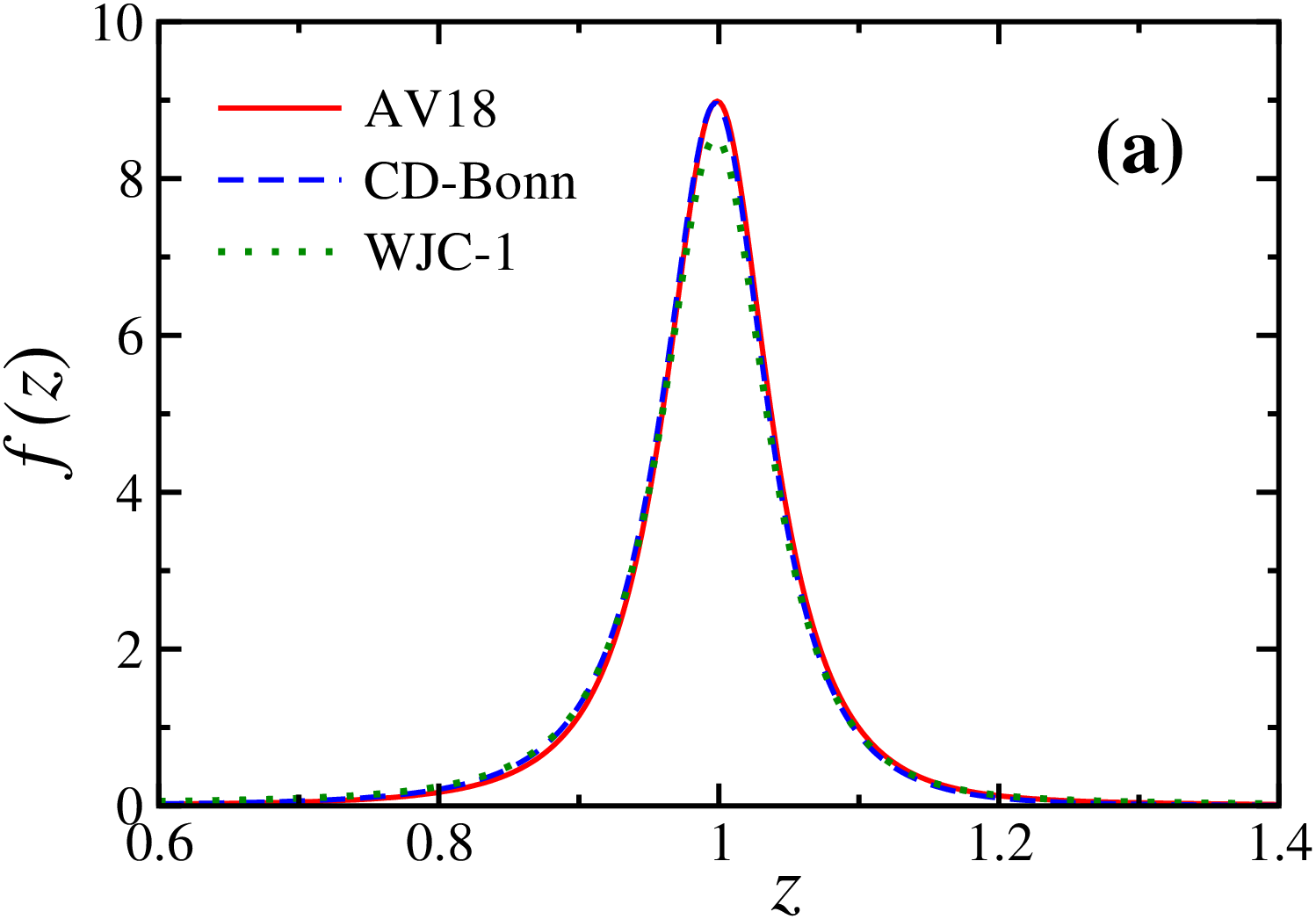}
\includegraphics[width=8.3cm]{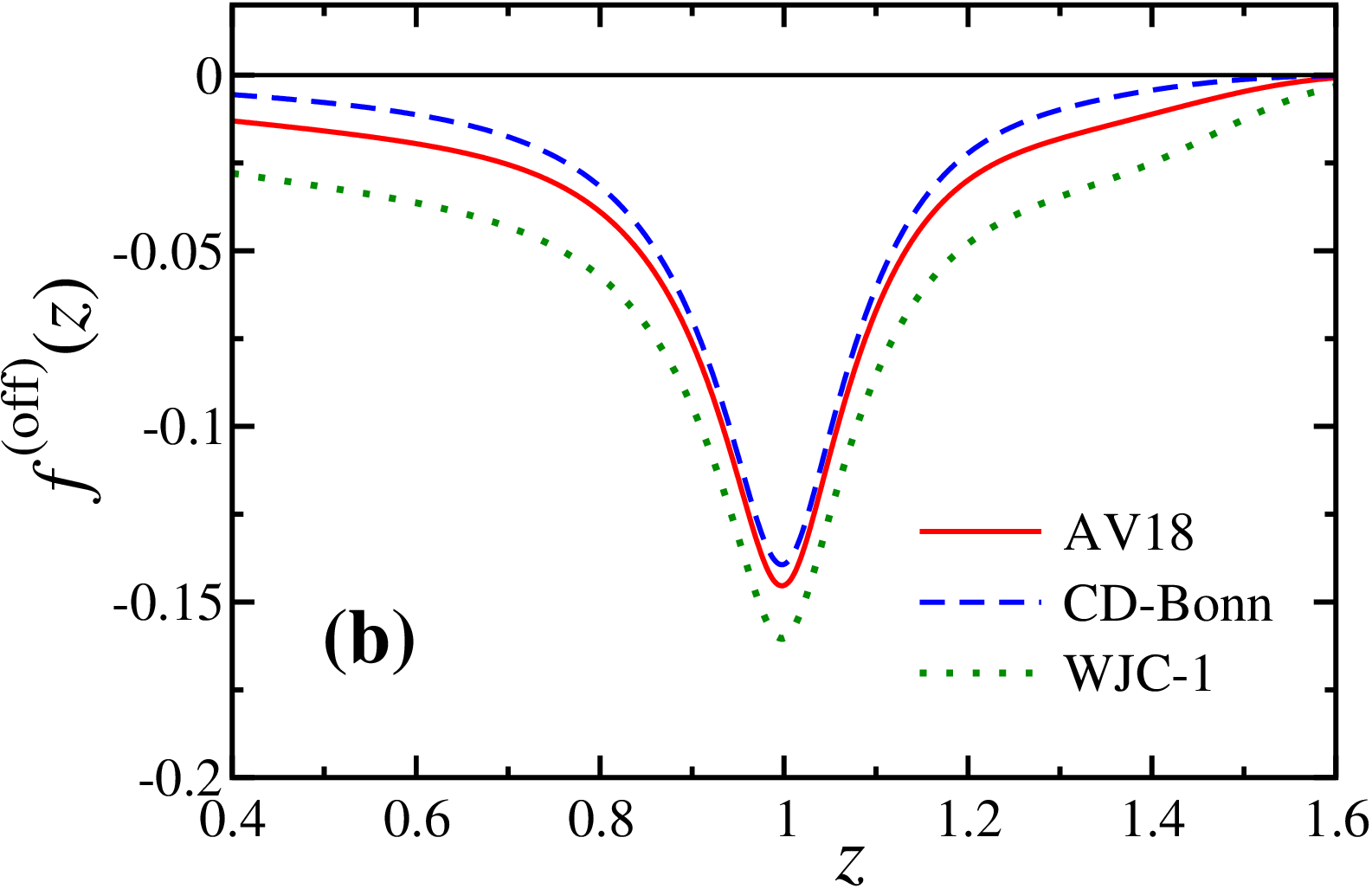}
\caption{Nucleon smearing functions in the deuteron:
	{\bf (a)} on-shell distribution $f(z)$ and
	{\bf (b)} off-shell function $f^{\rm off}(z)$,
	for the AV18 \cite{AV18} (solid lines),
	CD-Bonn \cite{CDBonn} (dashed lines), and
	WJC-1 \cite{WJC} (dotted lines)
	deuteron wave functions.}
\label{fig:fz}
\end{figure}

The $z$ dependence of the smearing function $f(z)$ and the off-shell
correction $f^{(\text{off})}(z)$ is illustrated in Fig.~\ref{fig:fz},
for several deuteron wave functions, based on the AV18 \cite{AV18},
CD-Bonn \cite{CDBonn} and WJC-1 \cite{WJC} nucleon--nucleon potentials.
As expected, the on-shell function $f(z)$ peaks strongly around
$z \approx 1$, and falls off rapidly away
from the peak.  The wave function dependence is relatively weak,
except at large $|z-1|$ \cite{Ethier14}.
Since the WJC-1 wave function has the hardest momentum distribution
of the models considered, the magnitude of the smearing function
is correspondingly smaller at the peak in order to preserve the
correct normalization.
In contrast, the off-shell smearing function $f^{(\text{off})}(z)$
displays a relatively stronger dependence on the deuteron wave
function, with the largest magnitude for the WJC-1 model and smallest
magnitude for the CD-Bonn wave function.  The negative sign of
$f^{(\text{off})}(z)$ arises from the $p^2-M^2$ factor in the
integrand in Eq.~(\ref{eq:foffz}), since the virtuality of the
off-shell nucleon is always less than $M^2$.  The small values of
the deuteron binding energy and average three-momentum distribution
suppress the magnitude of the off-shell function relative to $f(z)$
by about an order of magnitude.  However, as we shall see in the
next section, where we discuss the calculation of the $pN$ cross
section at the partonic level, the off-shell corrections can have
a significant effect on the overall $pd$ cross section.

\section{Parton level cross section}
\label{sec:pN}

To compute the $pd$ cross section in Eq.~(\ref{eq:pdxs_smeared})
requires calculation of the nucleon-level cross section in terms of
PDFs of the beam proton and bound nucleon in the target deuteron.
In this section we derive the proton--bound nucleon cross section,
working to leading order accuracy in the strong coupling.
After defining the kinematics relevant for the parton-level
process, we express the off-shell $pN$ cross section in terms
of off-shell generalizations of PDFs in the bound nucleon,
and construct a simple model to describe the possible $p^2$
dependence of the off-shell PDFs.

\subsection{Parton--parton scattering}
\label{ssec:parton}

In analogy with the definition of the external momentum variables
in the $pd$ frame in Eqs.~(\ref{eq:lcdecomp}), we decompose the
four-momenta of the colliding partons in the proton ($\hat k$)
and nucleon ($\hat p$) in terms of the light-cone vectors
$n^\mu$ and $\bar n^\mu$,
\begin{subequations}
\label{eq:partonic}
\begin{eqnarray}
\hat k^\mu
&=& \hat{k}^+\, \bar n^\mu
 +  \frac{\hat{k}^2 + \hat{\bm k}_\perp^2}{2 \hat{k}^+}\, n^\mu
 +  \hat{k}_\perp^\mu,			\\
\hat p^\mu
&=& \frac{\hat{p}^2 + \hat{\bm p}_\perp^2}{2 \hat{p}^-}\, \bar n^\mu
 +  \hat{p}^-\, n^\mu
 +  \hat{p}_\perp^\mu,
\end{eqnarray}
\end{subequations}
where $\hat{k}^2$ and $\hat{p}^2$ are the partons' virtualities
and $\hat{k}_\perp^\mu$ and $\hat{p}_\perp^\mu$ the respective
transverse momentum four-vectors.
In the collinear factorization framework, the total $pN$ amplitude
at leading twist is expressed as a product of the partonic hard
scattering amplitude and the soft, nonperturbative parton
distributions in the hadrons.
The partonic amplitude is calculated by expanding the parton momentum
about the direction of motion of the parent hadron and about the
parton's on-shell limit \cite{EFP83, Qiu90, AQ08}.
The hard scattering process can thus be computed by setting the
partonic momenta in Eqs.~(\ref{eq:partonic}) to
\begin{subequations}
\begin{eqnarray}
\label{eq:CF}
\hat{k}^+ &\to& \xi_p\, k^+,\ \ \ \
\hat{k}_\perp^\mu\ \to\ 0,		\\
\hat{p}^- &\to& \xi_N\, p^-,\ \ \ \,
\hat{p}_\perp^\mu\ \to\ \xi_N\, p_\perp^\mu,
\end{eqnarray}            
\end{subequations}  
which defines the partonic light-cone momentum fractions
$\xi_p$ and $\xi_N$ in the proton and target nucleon, respectively.
For light quarks, without loss of generality, one can take the
(on-shell) quarks to be massless, so that
\begin{eqnarray}
\hat{k}^2 \to 0,\ \ \ \ \hat{p}^2 \to 0.
\end{eqnarray}

At leading order in the strong coupling, the quark--antiquark pair
fuses into a virtual photon, which subsequently decays into a dilepton.
Conservation of four-momentum, $q^\mu = \hat{k}^\mu + \hat{p}^\mu$,
then implies that the momentum fractions are related by
\begin{equation}
  \xi_N\, =\, x_N\, =\, \frac{x_d}{z}, \ \ \ \ \
  \xi_p\, =\, \frac{Q^2}{Q_\perp^2}\, x_p,
\end{equation}
which can be obtained by equating the ``$-$'' and ``$+$'' components,
respectively, while from the transverse components one has
${\bm q}_\perp^2 = x_N^2\, {\bm p}_\perp^2$.
Since the average transverse momentum of the bound nucleon is
$\langle {\bm p}_\perp^2 \rangle \approx p_F^2 \sim 0.1$~GeV$^2$,
with $p_F$ the deuteron Fermi momentum, one can therefore neglect
the transverse momentum compared to the dilepton mass $Q^2$.
At high energies one then obtains $\xi_p \approx x_p$.
In this limit the proton--off-shell nucleon cross section becomes
\begin{eqnarray}
\label{eq:pN_LO} 
\widetilde{\sigma}^{pN}(x_p,x_N,p^2)
&=& \frac{4\pi\alpha^2}{9 x_p x_N s_N}
    \sum_q e_q^2
    \left[q(x_p) \widetilde{\bar q}(x_N,p^2)
	+ \bar q(x_p) \widetilde{q}(x_N,p^2)
    \right],
\end{eqnarray}
where $\widetilde{q}(x_N,p^2)$ is the PDF for a quark in
an off-shell nucleon with virtuality $p^2$, such that in
the on-shell limit, $p^2 \to M^2$, one has
$\widetilde{q}(x_N,M^2) \equiv q(x_N)$.
Note that at high energy $x_N s_N \approx x_d\, s$, so that
the dependence on $z$ enters only through the PDF arguments,
and the entire off-shell dependence of the cross section is
contained in the $p^2$ dependence of the PDFs.
In the next section we estimate this dependence in a simple
spectator model of the nucleon.

\subsection{Off-shell corrections}
\label{ssec:offshell}

In the absence of a first principles calculation of the nuclear
bound state in terms of quark and gluons degrees of freedom,
computing the off-shell behavior of PDFs is rather challenging.
Several attempts have been made in the literature to estimate the
$p^2$ dependence of the off-shell distribution $\widetilde{q}$
within effective quark models \cite{KP06, GL92, MSTplb, MSS97}.
Generally the models give rise to a suppression of the PDFs as
a function of $x$, although the quantitative features of the
off-shell modification depend somewhat on the details of the
specific model.

In this analysis we adopt the ``modified Kulagin-Petti'' model
\cite{KP06} for the valence distributions, used recently in the
CJ global PDF analyses \cite{CJ11, CJ12} which focused on the
high-$x$ region, and extend it to the sea quark sector to
describe the off-shell PDFs at both small and large $x$.
An attractive feature of this model is that the corrections can
be related to the average virtuality of the bound nucleons in the
nuclear medium, and the corresponding change of the nucleon's
confinement radius.  The valence PDFs for the bound nucleons
are further constrained by baryon number conservation, so that
the off-shell corrections do not alter the normalization.

In analogy with Eq.~(\ref{eq:sig_off}), we expand the off-shell PDF
$\widetilde{q}(x,p^2)$ to lowest order about its on-shell mass limit,
\begin{equation}
\label{eq:taylor}
\widetilde{q}(x,p^2)
= q(x) \left[1 + {(p^2-M^2) \over M^2}\, \delta q(x) \right],
\end{equation}
where the off-shell correction $\delta q(x)$ is given by
\begin{equation}
\label{eq:dltfq}
\delta q(x)
= {\frac{\partial\log \widetilde{q}}{\partial\log p^2}\vline}_{\,p^2=M^2}.
\end{equation}
Using Eqs.~(\ref{eq:pNxsgen}), (\ref{eq:sig_off}) and (\ref{eq:taylor}),
the off-shell correction to the proton--nucleon Drell-Yan cross section
can be written as
\begin{equation}
\label{eq:quark_expand}
\sigma^{pN}\, \delta\sigma^{pN}
= \frac{4\pi\alpha^2}{9 x_p x_N s_N}
  \sum_q e^2_q
  \left[ q(x_p)\, \bar{q}(x_N)\, \delta\bar{q}(x_N)\
      +\ \bar{q}(x_p)\, q(x_N)\, \delta q(x_N)
  \right],
\end{equation}
where the $(p^2-M^2)$ term in Eq.~(\ref{eq:sig_off}) has been
factored out.
At the parton level, the $pd$ differential cross section can then be
expressed (at leading order) in terms of the PDFs in the beam proton
and target deuteron,
\begin{eqnarray}
\sigma^{pd}(x_p,x_d)
&=& \frac{4\pi\alpha^2}{9 x_p\, x_d\, s_N}
    \sum_q e_q^2
    \left[ q(x_p)\, \bar q^{\, d}(x_d)\, +\, \bar q(x_p)\, q^d(x_d) \right],
\label{eq:sigma_qd}
\end{eqnarray}
where, in analogy with Eq.~(\ref{eq:pdxs_smeared}), the PDF in the
deuteron, $q^d$, is given by
\begin{eqnarray}
q^d(x_d)
&=& \sum_N \int_{x_d}^1 \frac{dz}{z}
    \left[ f(z) + f^{({\rm off})}(z)\, \delta q\Big({x_d\over z}\Big)
    \right]
    q^N\Big({x_d\over z}\Big),
\label{eq:q^d}
\end{eqnarray}
which includes corrections from nuclear smearing and nucleon off-shell
effects.

To evaluate the off-shell correction one assumes that the
$p^2$-dependent PDF can be represented in terms of a spectral
function $D_q$ \cite{KP06, KPW94},
\begin{equation}
\widetilde{q}(x,p^2)
= \int\!dw^2\int_{-\infty}^{\hat{p}_{\rm max}^2}d\hat{p}^2\,
  D_q(w^2,\hat{p}^2,x,p^2),
\label{eq:qoff}
\end{equation}
where $w^2=(p-\hat{p})^2$ is the mass of the (on-shell) spectator
quarks in the bound nucleon, and $\hat{p}^2$ is the interacting
quark's virtuality, with a maximum value
	$\hat{p}_{\rm max}^2 = x\, [p^2-w^2/(1-x)]$.
Note that since the PDF in Eq.~(\ref{eq:qoff}) represents the soft,
nonperturbative parton momentum distribution in the nucleon, there
is no hard scale to suppress contributions from off-shell partons,
in contrast to the hard scattering kinematics in Eqs.~(\ref{eq:CF}).
However, the spectral function $D_q$ must fall off sufficiently
fast at large $\hat{p}^2$ so as to ensure convergence of the
spectral integral.

Following Refs.~\cite{KPW94, KMPW95, KP06, CJ11}, we use the
single-pole approximation in which the spectator spectrum is
represented by a single (on-shell) state with effective mass
$\overline{w}_q^2$ for a given quark flavor $q$,
\begin{equation}
\label{eq:DqN}
D_q = \delta(w^2-\overline{w}_q^2)\,
      \Phi_q(\hat{p}^2,\Lambda(p^2)).
\end{equation}
Here the function $\Phi_q$ describes the momentum distribution
of quarks with virtuality $\hat{p}^2$ in the off-shell nucleon,
and the scale parameter $\Lambda(p^2)$ suppresses contributions
from large $\hat{p}^2$.
Note that in this approximation the $x$ dependence of the
off-shell distribution $\widetilde{q}$ arises from the upper
limit on the $\hat{p}^2$ integration in Eq.~(\ref{eq:qoff}),
which depends on $x$ as well as on $w^2$ and $p^2$.
In Ref.~\cite{KP06} the scale $\Lambda$ was related to the
confinement radius $R_N$ of the nucleon, $\Lambda \sim 1/R_N$,
which allows the $p^2$ dependence of $\Lambda$ to be interpreted
in terms of the change in the nucleon radius when the nucleon is
bound inside the deuteron.
Using Eq.~(\ref{eq:DqN}), the off-shell correction $\delta q$
in Eq.~(\ref{eq:taylor}) can then be written
\begin{eqnarray}
\label{eq:fq}
\delta q(x)
&=& C_q + \frac{\partial\log{q}}{\partial x} h_q(x),
\end{eqnarray}
where $C_q$ is determined by PDF normalization constraints, and
\begin{eqnarray}
h_q(x)
&=& x (1-x) \frac{(1-\lambda)(1-x) + \lambda\overline{w}_q^2/M^2}
		 {(1-x)^2 - \overline{w}_q^2/M^2},
\end{eqnarray}
with the parameter $\lambda$ defined as
\begin{eqnarray}
\lambda
&=& {\frac{\partial\log\Lambda^2}
	  {\partial\log p^2}\vline}_{\,p^2=M^2}.
\end{eqnarray}
Since the cutoff $\Lambda$ is inversely proportional to $R_N$,
one can also write the $\lambda$ parameter as
	$\lambda = -2 (\delta R_N/R_N) (\delta p^2/M^2)$,
where $\delta R_N$ is the change in the nucleon's radius in
the nuclear medium, and
	$\delta p^2/M^2 = \int dz\, f^{(\rm off)}(z)$
is the average nucleon virtuality in the deuteron.
The value of $\delta p^2$ depends on the $NN$ potential model,
and ranges between $\delta p^2/M^2 \approx -3.6\%$ and $-6.5\%$
for the \mbox{CD-Bonn} \cite{CDBonn} and WJC-1 \cite{WJC}
deuteron wave functions, respectively, with other wave functions
such as Paris \cite{Paris} and AV18 \cite{AV18} giving
intermediate values.
Estimates of the change in confinement radius in the deuteron
based on the analysis of data on the nuclear EMC effect suggest
\cite{CJRR} a value for $\delta R_N/R_N \sim {\cal O}(1\%-2\%)$.
In Ref.~\cite{CJ12} the uncertainties in the off-shell corrections
were estimated by considering several combinations of the deuteron
wave function and the nucleon ``swelling'', and were represented in
the form of the ``CJ12min'' (small nuclear correction), ``CJ12mid''
(medium nuclear correction) and ``CJ12max'' (large correction) PDFs.
These correspond, respectively, to the WJC-1 wave function (hardest)
with a 0.3\% nucleon swelling, the AV18 wave function with a 1.2\%
swelling effect, and the CD-Bonn wave function (softest) with a
2.1\% nucleon swelling.

In previous calculations of off-shell corrections to PDFs using this
type of model \cite{KPW94, KMPW95, KP06, CJ11} (which, following
Ref.~\cite{Ethier13}, we refer to generally as the ``off-shell
covariant spectator'' or OCS model), only valence quark distributions
were studied.
However, with reasonable approximations it is straightforward
to extend this model to compute the off-shell corrections to
sea quark (and gluon) distributions, as needed in the analysis
of Drell-Yan data.
(We include gluons here for completeness, even though the gluon PDF
does not enter explicitly in the leading order Drell-Yan cross section;
it will be relevant, however, for next-to-leading order calculations.)
The essential difference will be in the values of the spectator
effective mass $\overline{w}_q^2$ for a given parton flavor.
Within the OCS model framework, a fit to existing PDFs in the
free proton gives
$\overline{w}^2_v = 2.2$~GeV$^2$,
$\overline{w}^2_s = 5.5$~GeV$^2$, and
$\overline{w}^2_g = 8.0$~GeV$^2$
for the valence, sea, and gluon distributions, respectively.
The valence mass parameter is similar to that found in Ref.~\cite{KP06},
and the larger masses for the sea quark and gluon distributions reflect
the larger minimum number of partons required in the intermediate state
for sea quarks and gluons compared to valence quarks.

The normalization constant $C_q$ in Eq.~(\ref{eq:fq}) is computed for
valence quarks $q_v = q - \bar q$ by requiring that the off-shell
correction does not alter the baryon number \cite{CJ11},
\begin{eqnarray}
\int_0^1 dx\, q_v(x)\, \delta q_v(x) &=& 0.
\end{eqnarray}
For sea quarks and gluons, on the other hand, perturbative radiation
of soft gluons and generation of $q\bar q$ pairs render the
corresponding integrals infinite.  To proceed one can either impose
a constraint from a higher moment, such as the momentum sum rule,
or alternatively consider a model of the nucleon at a low momentum
scale involving a finite number of partonic degrees of freedom.
Parametrizations based on this boundary condition have long been
utilized by the Dortmund group \cite{GRV98, JR09, JR14}, for example,
assuming valence-like gluon and sea PDFs at low $Q^2$, and generating
the high-$Q^2$ dependence through perturbative evolution.
This in fact is closer in spirit to the spectral function model
of Eq.~(\ref{eq:qoff}) with finite values of the spectator system
mass $\overline{w}_q^2$.

We considered both methods of determining the sea normalization,
but found relatively small differences at medium to high values
of $x$ ($x \gtrsim 0.3$).  At smaller $x$, results from Drell-Yan
measurements of cross section ratios of C, Ca, Fe and W to deuterium
\cite{Alde90, McGaughey92} in the Fermilab E772 experiment disfavor
large medium modifications of antiquark distributions for
$0.1 \lesssim x \lesssim 0.3$.
In this range, the OCS model with the momentum sum rule constraint
gives a fairly small correction, albeit with sizeable uncertainty,
and to ensure consistency with the E772 Drell-Yan data we therefore
smoothly extrapolate the corrections to zero below $x \approx 0.15$.

\begin{figure}[t]
\includegraphics[width=10cm]{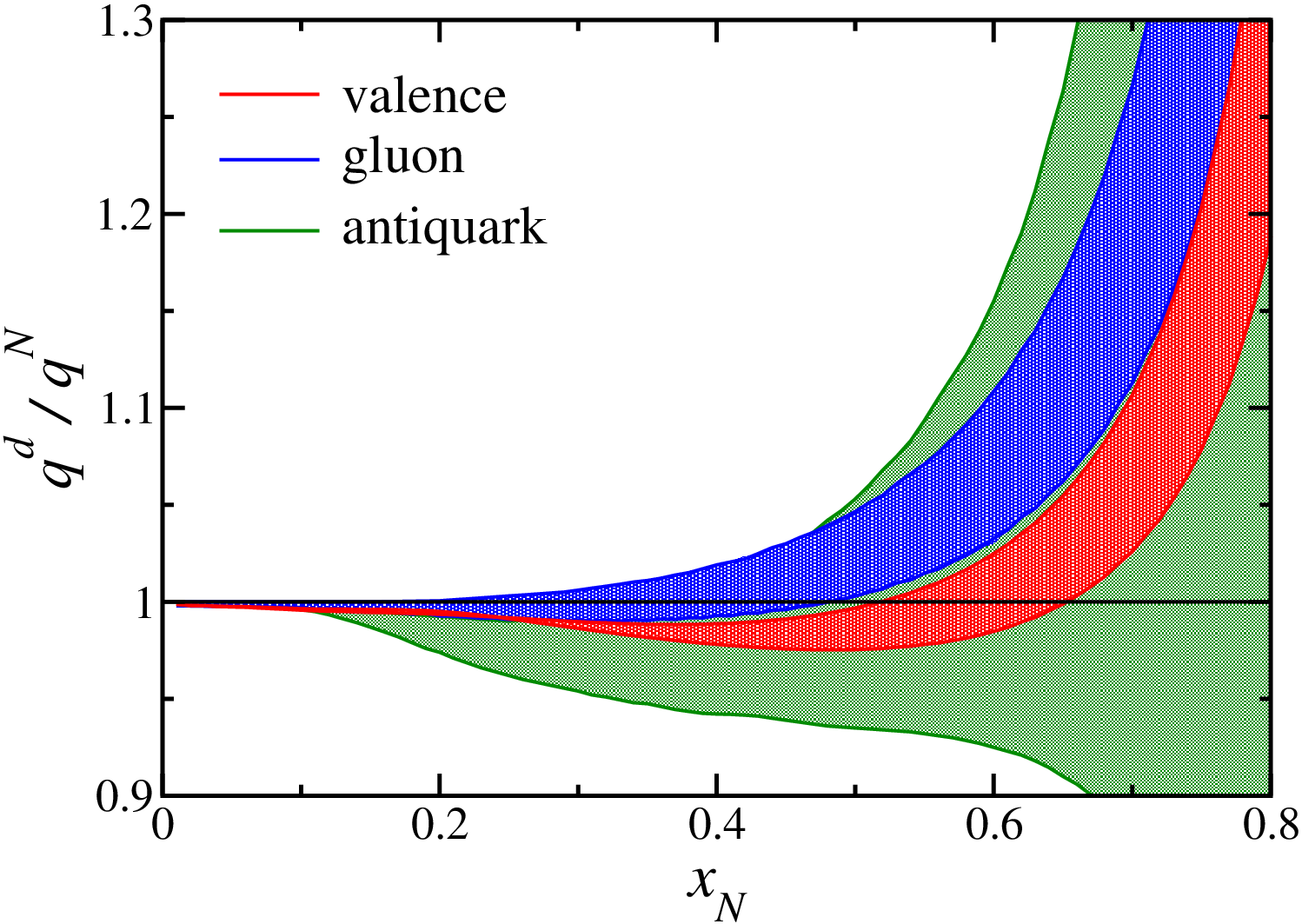}
\caption{Ratio of PDFs in the deuteron to those in an isoscalar nucleon,
	$q^d/q^N$, for valence $q=u_v+d_v$ quarks (red shaded band),
	antiquarks $q=\bar u+\bar d$ (green shaded band), and gluons
	(blue shaded band), including the effects of nuclear smearing
	and nucleon off-shell corrections.
	The bands represent the range of deuteron wave functions
	and nucleon off-shell parameters used in Ref.~\cite{CJ12}.}
\label{fig:qoff}
\end{figure}

The ratios $q^d/q^N$ of the PDFs in the deuteron, calculated
through Eq.~(\ref{eq:q^d}), to those in an isoscalar nucleon
($N=p+n$) are illustrated in Fig.~\ref{fig:qoff} for the valence
quark $u_v+d_v$, sea quark (or antiquark) $\bar u+\bar d$,
and gluon distributions.
For consistency with the original analysis of the Drell-Yan
data from the E866 experiment \cite{E866}, we use the input
nucleon PDFs from the CTEQ5 global QCD analysis \cite{CTEQ5}
as in Ref.~\cite{E866}.
The distributions in the deuteron include the effects of nuclear
smearing and nucleon off-shell corrections, with the bands in
Fig.~\ref{fig:qoff} illustrating the maximal range from different
models of the deuteron wave function and nucleon ``swelling''
parameters, as discussed above.
Specifically, the upper edges of the bands, with the largest
$q^d/q^N$ ratios, correspond to the smallest nuclear corrections
(WJC-1 deuteron wave function with a 0.3\% nucleon swelling),
while the lower edges correspond to stronger nuclear corrections
(CD-Bonn wave function and up to $\sim 2\%$ nucleon swelling).

The effects of the nuclear smearing are evident in the rise above
unity at $x \gtrsim 0.5$ of the valence quarks and gluon ratios,
characteristic of nuclear deep-inelastic structure function ratios
in the nuclear EMC effect \cite{CJ12}.
The behavior of the antiquark ratios for the smallest nuclear
corrections is similar at intermediate and large $x$, while for
the strongest nuclear corrections the ratio stays below unity
over the range $x \lesssim 0.7$ over which the antiquark PDFs
are determined in the CTEQ5 fit \cite{CTEQ5}.
The large spread in the antiquark ratios for $x \gtrsim 0.4$
results from a combined effect of the nuclear correction
uncertainties, and the very small size (with large uncertainty)
of the $\bar u$ and $\bar d$ PDFs in this region.

Finally, we also note that the general form of the convolution
in Eqs.~(\ref{eq:sigma_qd}) and (\ref{eq:q^d}) closely resembles
the result obtained recently by Kamano and Lee \cite{KL12},
although with some important differences.
In particular, whereas the momentum fraction $z$ here is defined
in the context of collinear factorization on the light-cone
\cite{Collins88}, as a fraction of the ``minus'' components of
the proton and deuteron four-vectors [Eq.~(\ref{eq:z})], in
Ref.~\cite{KL12} it is related to the ratio $p_z/p_z^{\rm ave}$
of the nucleon's longitudinal momentum relative to its quadratic
average with respect to the deuteron wave function,
$p_z^{\rm ave} = \langle p_z^2 \rangle^{1/2}$.
Therefore, although the formal expressions for the $pd$ cross sections
are similar, a direct comparison of the cross sections derived here
and in Ref.~\cite{KL12} is difficult because of the different frames
and variables used in the two approaches.
An advantage of the present approach is that by working with light-cone
momentum fractions our results are invariant under Lorentz boosts along
the light-front.  In practice we perform the calculation in the rest
frame of the deuteron, so that no approximations need to be made in
boosting the deuteron wave function.

\newpage
\section{Nuclear effects on cross section ratios}
\label{sec:results}

\begin{figure}[t]
\includegraphics[width=10cm]{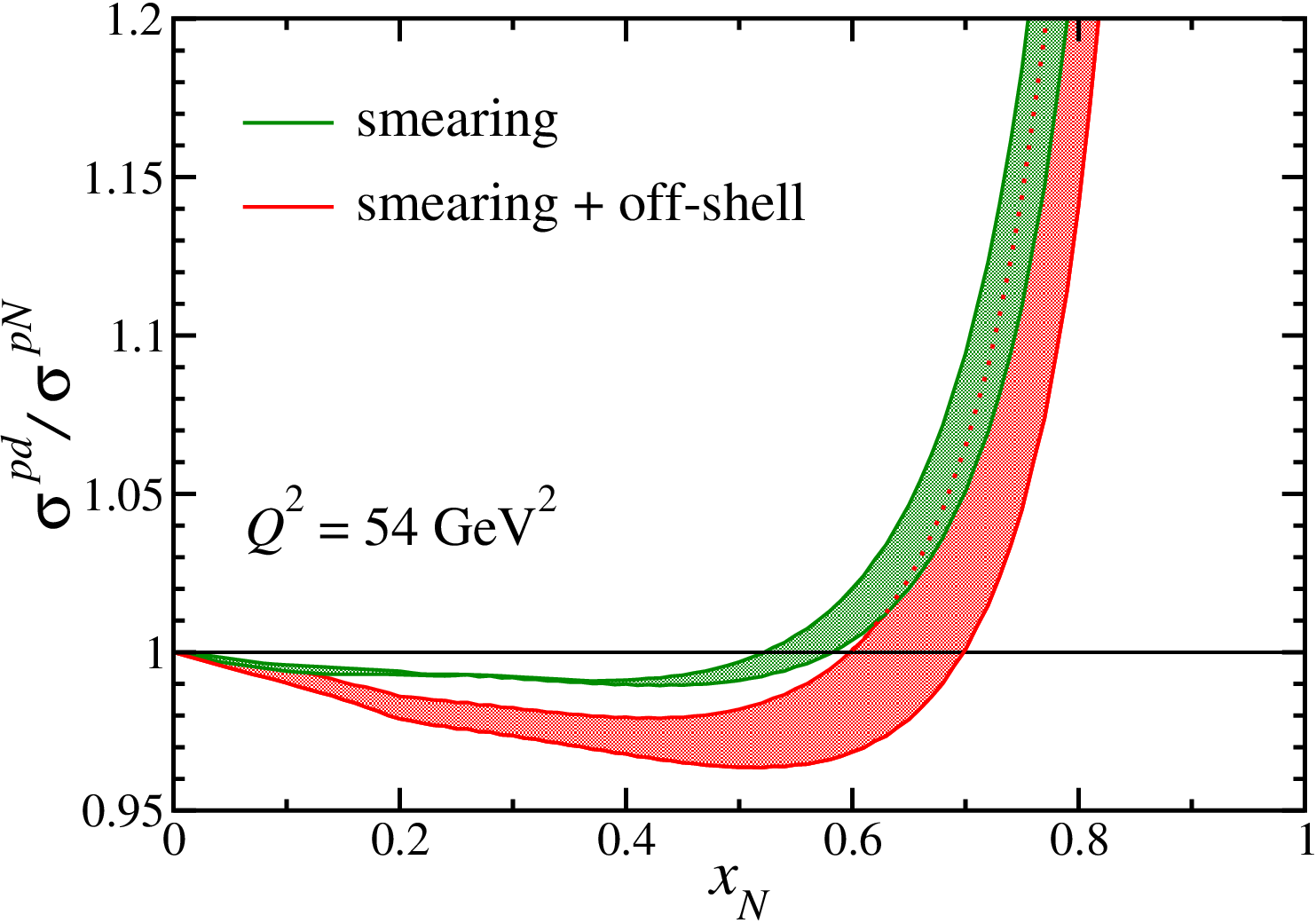}
\caption{Ratio of $pd$ to $pN$ Drell-Yan cross sections 
	$\sigma^{pd}/\sigma^{pN}$ at the kinematics corresponding
	to the Fermilab E866 experiment \cite{E866}
	(incident proton energy $k_0 = 800$~GeV and average
	$Q^2=54$~GeV$^2$), including the effects of nuclear smearing
	(green band) and smearing $+$ nucleon off-shell corrections.
	The band for the smearing only corrections represents the
	range defined by the WJC-1 \cite{WJC}, AV18 \cite{AV18}	and
	CD-Bonn \cite{CDBonn} deuteron wave functions, while the
	smearing $+$ off-shell band includes in addition the range
	of nucleon off-shell (swelling) parameters \cite{CJ12}.}
\label{fig:pd_pN}
\end{figure}

Using the formalism for the $pd$ Drell-Yan cross section derived in
Sec.~\ref{sec:pd} and the OCS model for the off-shell nucleon PDFs
in the deuteron in Sec.~\ref{sec:pN}, we can proceed to compute the
nuclear effects in the $pd$ dilepton production reaction by comparing
them with the corresponding proton--free nucleon scattering process.
In Fig.~\ref{fig:pd_pN} we illustrate the effects of the nuclear
corrections in the ratio of the $pd$ to the isoscalar nucleon $pN$
Drell-Yan cross section, computed at the kinematics of the Fermilab
E866 \cite{E866} experiment (incident energy $k_0=800$~GeV and
average $Q^2=54$~GeV$^2$).
The ratio displays the characteristic shape of the nuclear EMC effect,
with a small, few percent depletion at intermediate values of $x$
($x \lesssim 0.5$) and a rapid rise above unity at larger $x$
($x \gtrsim 0.6$).  The greater spread in the calculated ratio
in the high-$x$ region ($x \gtrsim 0.5$) reflects the larger
uncertainties in the deuteron wave function at short distances,
or large $z$ in the nuclear smearing function in Fig.~\ref{fig:fz}.
The nucleon off-shell corrections act to reduce the $pd$ cross section
over most of the range of $x$, resulting in a more sizeable nuclear
effect at intermediate $x$ ($0.2 \lesssim 0.6$).  The overall
uncertainty also increases due to the range of possible behaviors
of the bound nucleon PDFs, as discussed in Sec.~\ref{sec:pN}.

\begin{figure}[t]
\includegraphics[width=10cm]{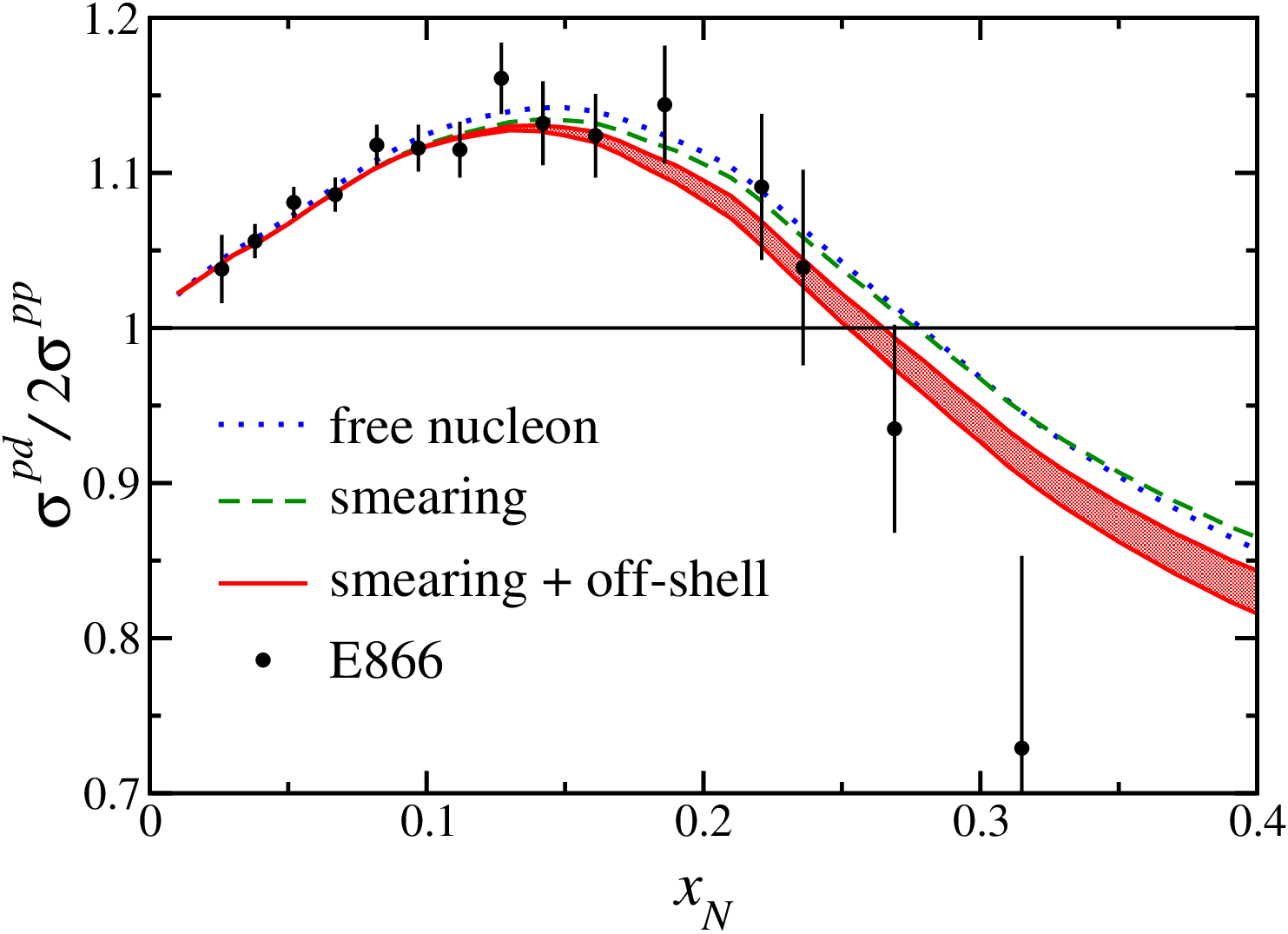}
\caption{Ratio of $pd$ to $pp$ Drell-Yan cross sections
	$\sigma^{pd}/2\sigma^{pp}$ at the kinematics of the
	Fermilab E866 data \cite{E866}, with incident proton
	energy $k_0 = 800$~GeV and average $Q^2=54$~GeV$^2$.
	The E866 data (filled circles) are compared with the
	free nucleon calculation without any nuclear effects
	(blue dotted curve), with nuclear smearing corrections
	only (green dashed curve), and with nuclear smearing
	$+$ off-shell corrections (red solid band).}
\label{fig:e866}
\end{figure}

The reduction of the $pd$ Drell-Yan cross section in the presence
of nuclear corrections, relative to the $pN$ cross section, is
clearly visible in the $pd$ to $pp$ cross section ratio shown
in Fig.~\ref{fig:e866}.  While the effect of the nuclear smearing
is relatively mild over the range of $x$ covered by the Fermilab
E866 data, $0.02 \lesssim x \lesssim 0.3$
(qualitatively similar to that found in Ref.~\cite{KL12}),
the addition of nucleon off-shell corrections lowers the overall
free-nucleon cross section appreciably for $x \gtrsim 0.1-0.2$.
An intriguing feature of the E866 data is the apparent reduction of
the $pd$ to $pp$ cross section ratio below unity at the two largest-$x$
data points, albeit with large uncertainties, suggesting a possible
sign change of $\bar d-\bar u$ for $x \gtrsim 0.25$.
By lowering the $pd$ cross sections in this region, the nuclear
corrections computed here improve the agreement with the data in
this region (using the CTEQ5 PDF set \cite{CTEQ5}), although it is
unlikely that this can account for the entire effect at large $x$.

\begin{figure}[t]
\includegraphics[width=10cm]{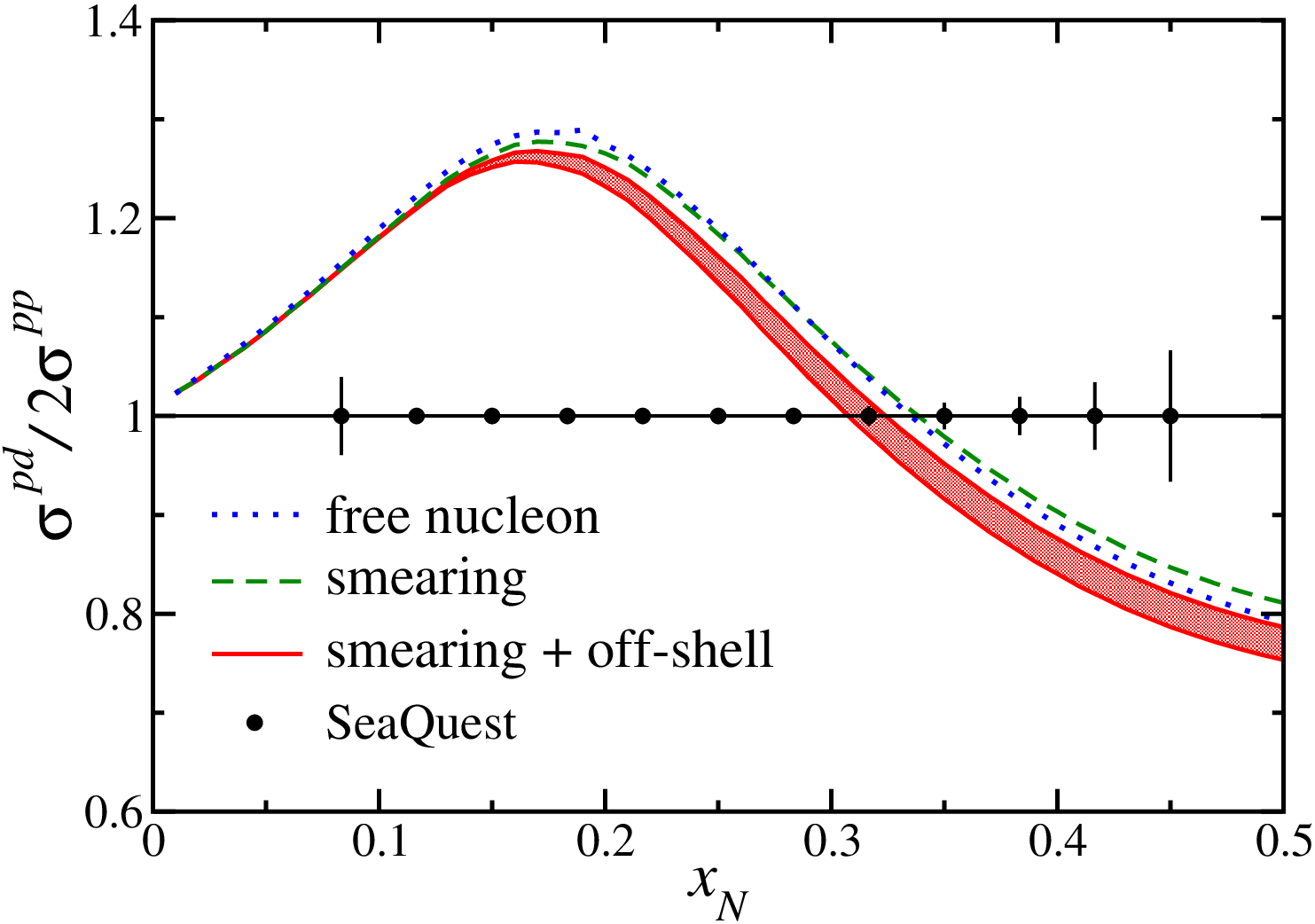}
\caption{Ratio of $pd$ to $pp$ Drell-Yan cross sections as in
	Fig.~\ref{fig:e866}, but at the kinematics of the
	new Fermilab SeaQuest experiment \cite{E906},
	with incident proton energy $k_0 = 120$~GeV and average
	$Q^2=42$~GeV$^2$.  The projected data (filled circles)
	with estimated uncertainties are arbitrarily placed
	at unity.}
\label{fig:e906}
\end{figure}

To better understand the large-$x$ behavior of $\bar d/\bar u$,
the new Fermilab SeaQuest experiment (``SeaQuest''), using a lower
proton beam energy of $k_0 = 120$~GeV, was proposed to measure the
$pd$ to $pp$ Drell-Yan cross section ratio to $x \approx 0.45$
\cite{E906}.  With the expected precision of the data illustrated
in Fig.~\ref{fig:e906}, this measurement should unambiguously
determine the trend of the $\bar d/\bar u$ ratio in the
$x \approx 0.3-0.4$ region.
Computing the nuclear effects on the $pd$ cross section at the
SeaQuest kinematics, the impact of the nuclear smearing and
off-shell corrections is comparable to or even larger than
the projected uncertainties for $0.15 \lesssim x \lesssim 0.4$.
This suggests that the overall systematic uncertainties in the
measurement may be underestimated in this region, and that further
work may be needed in order to better constrain the theoretical
uncertainties in the calculation of the nuclear corrections to
the $pd$ cross section.
The SeaQuest experiment commenced data taking in 2014, and is
expected to run until late 2015, with first results anticipated
by the end of 2014 \cite{Peng-priv}.

\begin{figure}[t]
\includegraphics[width=10cm]{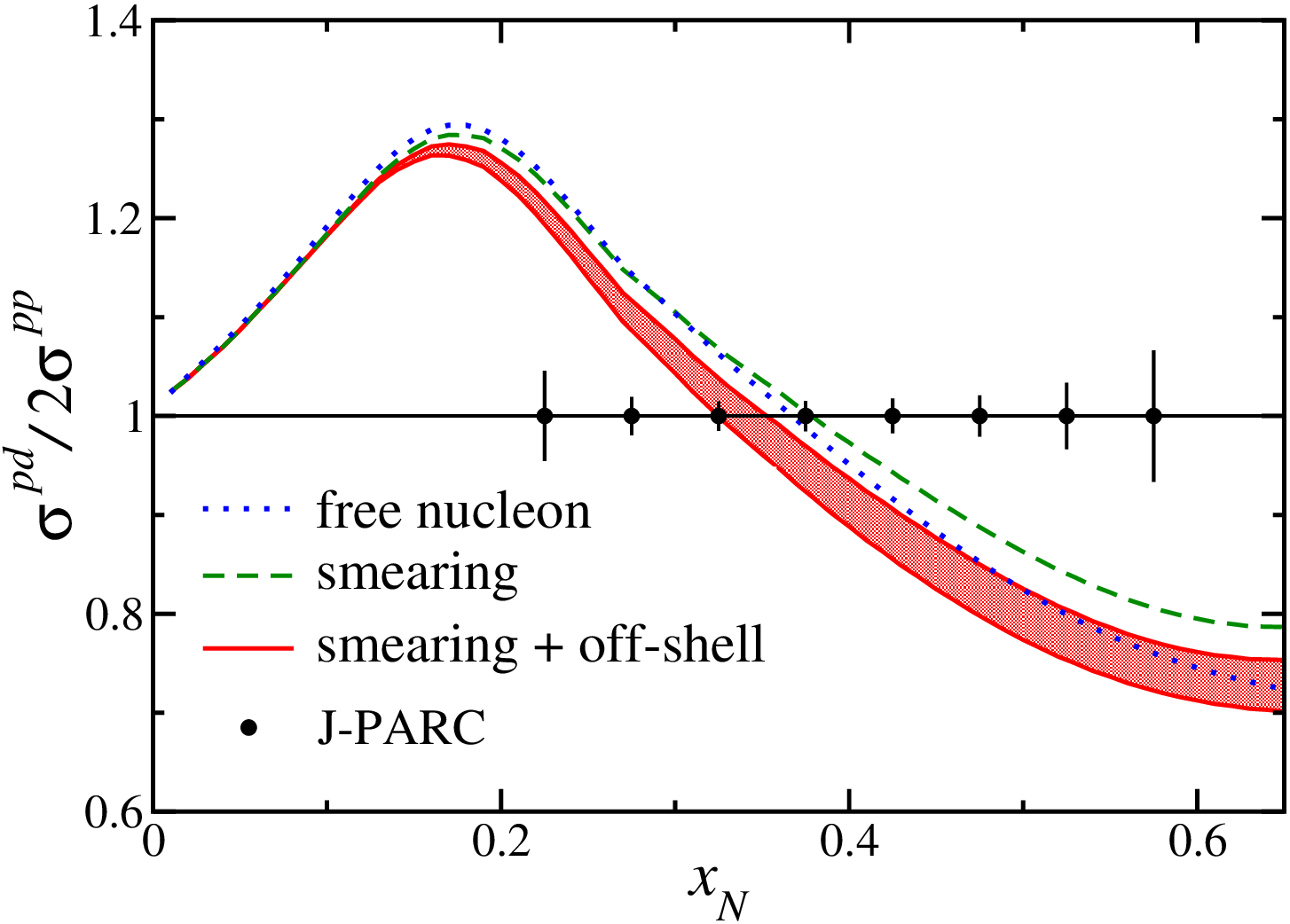}
\caption{Ratio of $pd$ to $pp$ Drell-Yan cross sections as in
	Fig.~\ref{fig:e866}, but at the kinematics of the proposed
	J-PARC experiment \cite{J-PARC}, with incident proton
	energy $k_0 = 50$~GeV and average $Q^2 = 25$~GeV$^2$.
	The projected data (filled circles) with estimated
	uncertainties are arbitrarily placed at unity.}
\label{fig:j-parc}
\end{figure}

Beyond the Fermilab experiments, a proposal has been made to extend
the Drell-Yan cross section measurements to even larger $x$
($x \lesssim 0.6$) at the J-PARC facility in Japan \cite{J-PARC},
using a 50~GeV proton beam.  The size of the nuclear smearing and
nucleon off-shell corrections to the $pd/pp$ cross section ratio is
illustrated in Fig.~\ref{fig:j-parc} for an average dilepton mass
of $Q^2 = 25$~GeV$^2$, together with the expected experimental
uncertainties.  Because of the higher values of $x$ that would be
probed here, the effects of the nuclear smearing are expected to
be correspondingly more significant than for the lower-$x$ data
from the E866 and SeaQuest experiments, as can be anticipated from
Fig.~\ref{fig:pd_pN}.  As for the SeaQuest data, the uncertainty
range of the nuclear corrections is similar to or larger than
the projected experimental uncertainties for
$0.25 \lesssim x \lesssim 0.5$, again suggesting the need to
better constrain the nuclear corrections if the planned precision
is to be reached.
Currently, the J-PARC facility is approved for 30~GeV proton running;
an upgrade to a 50~GeV proton beam would be needed to realize the
goals of the proposed experiment \cite{J-PARC, Peng-priv}.

\section{Conclusions}
\label{sec:conc}

With the significant improvement in the determination of the
$\bar d/\bar u$ ratio at large values of $x$ expected from upcoming
experiments, particularly the SeaQuest Drell-Yan experiment at
Fermilab \cite{E906}, preliminary results from which are
anticipated in 2015 \cite{Peng-priv}, the need exists to
understand the computation of the cross section with sufficient
accuracy for an unambiguous extraction of the signal.
In this study we have derived the proton--deuteron dilepton
production cross section in terms of the proton--nucleon
cross section, paying particular attention to nuclear
smearing and nucleon off-shell corrections in the deuteron.
The form of the relation between the nuclear and nucleon level
cross sections resembles the familiar convolution result from
deep-inelastic scattering \cite{AKL04, KP06, Kahn09, AQV}:
in the high energy limit the nucleon light-cone momentum distribution
in the deuteron in $pd$ scattering is found to correspond exactly
to the Bjorken limit smearing function relevant for describing
electron--deuteron scattering in the weak binding approximation
\cite{60fest}.

The effects of Fermi motion and nuclear binding contained in the
nuclear smearing function are relatively small in the region of $x$
spanned by the existing E866 data, but become more noticeable at the
higher $x$ values ($x \gtrsim 0.4$) that will be accessed in the new
Fermilab \cite{E906} and proposed J-PARC \cite{J-PARC} experiments.
Corrections arising from the possible off-shell deformation of the
nucleon PDFs in the deuteron have been estimated within a simple
spectator model of the nucleon that has previously been applied
to the analysis of lepton--deuteron deep-inelastic data
\cite{KP06, CJ11, CJ12}.
For the same range of nucleon swelling parameters as those adopted
in the recent CJ global PDF analysis \cite{CJ12}, the ratio of
$pd$ to $pp$ Drell-Yan cross sections is found to be significantly
reduced compared with the free-nucleon calculation.
Furthermore, while the nuclear model uncertainty, from both the
short-distance part of the deuteron wave function and the nucleon
off-shell parameters, is small compared with the existing E866 data,
it is of similar size to or even larger than the projected
uncertainties for the new SeaQuest experiment in the range
$0.15 \lesssim x \lesssim 0.4$.  Generally, the magnitude of the
corrections and their uncertainties increase with increasing 
values of $x$.

These findings suggest that the overall systematic uncertainties
in the future measurements may be underestimated at large $x$,
and that further work may be warranted to reduce the theoretical
uncertainties on the $pd$ cross section in order to attain the
precision goal of the experiments.
Although the exact magnitude of the nuclear corrections is subject
to some model dependence, the sign of the effect appears universal.
In particular, the reduction of the $pd$ cross sections will lead to
an increased $\bar d/\bar u$ ratio extracted from global PDF analyses,
with the largest effects expected at the highest $x$ values.
While in the present work we have made use of the CTEQ5 parametrization
of global PDFs \cite{CTEQ5} to illustrate the systematics of the
nuclear corrections, the results from this analysis will be used in
future global QCD fits \cite{CJ14} to obtain a more reliable estimate
of the light quark sea distributions in the proton.
For future theoretical work, it will also be necessary to reexamine
the pion-exchange corrections to $pd$ scattering, which were found
in Ref.~\cite{KL12} to be significant at large $x$.
Earlier work on pion-exchange in lepton--deuteron deep-inelastic
scattering \cite{Kaptari91, MT93, Nikolaev97} suggested that
pion-exchange corrections were important primarily at lower
$x$ values, $x \sim 0.1$.
A combined analysis of both nucleonic and pionic contributions
within our collinear framework, as well as nuclear shadowing
corrections at small $x$, would then enable a complete description
of the nuclear effects in the $pd$ Drell-Yan reaction.

\acknowledgments

We thank D.~F.~Geesaman, S.~Kumano and J.-C.~Peng for helpful
communications about the proposed Drell-Yan experiments at
Fermilab and J-PARC.
This work was supported by the DOE Contract No.~DE-AC05-06OR23177,
under which Jefferson Science Associates, LLC operates Jefferson Lab,
and by the NSF and DOD's ASSURE program.
The work of A.A. was supported in part by DOE Contract
No.~DE-SC0008791.



\begin{thebibliography}{99}

\bibitem{NMC91}
P.~Amaudruz {\it et al.},
Phys. Rev. Lett. {\bf 66}, 2712 (1991).

\bibitem{NMC94} 
M.~Arneodo {\it et al.},
Phys. Rev. D {\bf 50}, 1 (1994).

\bibitem{HERMES}
K.~Ackerstaff {\it et al.},
Phys. Rev. Lett. {\bf 81}, 5519 (1998).

\bibitem{NA51}
A.~Baldit {\it et al.},
Phys. Lett. B {\bf 332}, 244 (1994).

\bibitem{E866_98}
E.~A.~Hawker {\it et al.},
Phys. Rev. Lett. {\bf 80}, 3715 (1998).

\bibitem{E866}
R.~S.~Towell {\it et al.},
Phys. Rev. D {\bf 64}, 052002 (2001).

\bibitem{Kumano98}
S.~Kumano,
Phys. Rep. {\bf 303}, 183 (1998).

\bibitem{Speth98}
J.~Speth and A.~W.~Thomas,
Adv. Nucl. Phys. {\bf 24}, 83 (1998).

\bibitem{Garvey01}
G.~T.~Garvey and J.-C.~Peng,
Prog. Part. Nuc. Phys. {\bf 47}, 203 (2001).

\bibitem{Peng14}
J.-C.~Peng and J.-W.~Qiu,
Prog. Part. Nucl. Phys. {\bf 76}, 43 (2014).

\bibitem{DY70}
S.~D.~Drell and T.~M.~Yan,
Phys. Rev. Lett. {\bf 25}, 316 (1970).

\bibitem{Ellis91}
S.~D.~Ellis and W.~J.~Stirling,
Phys. Lett. B {\bf 256}, 258 (1991).

\bibitem{AKL04}
S.~I.~Alekhin, S.~A.~Kulagin and S.~Liuti,
Phys. Rev. D {\bf 69}, 114009 (2004).

\bibitem{KP06}
S.~A.~Kulagin and R.~Petti,
Nucl. Phys. {\bf A765}, 126 (2006).

\bibitem{Kahn09}
Y.~Kahn, W.~Melnitchouk and S.~Kulagin,
Phys. Rev. C {\bf 79}, 035205 (2009).

\bibitem{AQV}
A.~Accardi, J.~W.~Qiu, and J.~P.~Vary,
``Collinear factorization and deep inelastic scattering
on nuclear targets'' (2011, unpublished).       

\bibitem{ABKM09}
S.~Alekhin, J.~Bl\"umlein, S.~Klein and S.-O.~Moch,
Phys. Rev. D {\bf 81}, 014032 (2010).

\bibitem{CJ10}
A.~Accardi, M.~E.~Christy, C.~E.~Keppel, P.~Monaghan, W.~Melnitchouk,
J.~G.~Morfin and J.~F.~Owens,
Phys. Rev. D {\bf 81}, 034016 (2010).

\bibitem{CJ11}
A.~Accardi, W.~Melnitchouk, J.~F.~Owens, M.~E.~Christy, C.~E.~Keppel,
L.~Zhu, and J.~G.~Morfin, 
Phys. Rev. D {\bf 84}, 014008 (2011).

\bibitem{CJ12}
J.~F.~Owens, A.~Accardi and W.~Melnitchouk,
Phys. Rev. D {\bf 87}, 094012 (2013).

\bibitem{Brady12}
L.~T.~Brady, A.~Accardi, W.~Melnitchouk and J.~F.~Owens,
JHEP {\bf 1206}, 019 (2012).

\bibitem{JMO13}
P.~Jimenez-Delgado, W.~Melnitchouk and J.~F.~Owens,
J. Phys. G: Nucl. Part. Phys. {\bf 40}, 093102 (2013).    

\bibitem{MST98} 
W.~Melnitchouk, J.~Speth and A.~W.~Thomas,
Phys. Rev. D {\bf 59}, 014033 (1998).

\bibitem{KL12}
H.~Kamano and T.~-S.~H.~Lee,
Phys. Rev. D {\bf 86}, 094037 (2012).

\bibitem{E906}
Fermilab E906 experiment (SeaQuest),
{\it Drell-Yan Measurements of Nucleon and Nuclear Structure with the
Fermilab Main Injector},
D.~F.~Geesaman and P.~E.~Reimer, spokespersons;
{\tt http://www.phy.anl.gov/mep/SeaQuest/index.html}.

\bibitem{Kaptari91}
L.~P.~Kaptari and A.~Yu.~Umnikov,
Phys. Lett. B {\bf 272}, 359 (1991).

\bibitem{MT93}
W.~Melnitchouk and A.~W.~Thomas,
Phys. Rev. D {\bf 47}, 3783 (1993).

\bibitem{Nikolaev97}
N.~N.~Nikolaev and W.~Schafer,
Phys. Lett. B {\bf 398}, 245 (1997),
[Erratum-ibid. B {\bf 407}, 453 (1997)].

\bibitem{GL92}
F.~Gross and S.~Liuti,
Phys. Rev. C {\bf 45}, 1374 (1992).

\bibitem{Ciof92}
C.~Ciofi degli Atti, D.~B.~Day and S.~Liuti,
Phys. Rev. C {\bf 46}, 1045 (1992).

\bibitem{MST94}
W.~Melnitchouk, A.~W.~Schreiber and A.~W.~Thomas,
Phys. Rev. D {\bf 49}, 1183 (1994).

\bibitem{MSTplb}
W.~Melnitchouk, A.~W.~Schreiber and A.~W.~Thomas,
Phys. Lett. B {\bf 335}, 11 (1994).

\bibitem{KPW94}
S.~A.~Kulagin, G.~Piller and W.~Weise,
Phys. Rev. C {\bf 50}, 1154 (1994).

\bibitem{Cosyn13}
W.~Cosyn, W.~Melnitchouk and M.~Sargsian,
Phys. Rev. C {\bf 89}, 014612 (2014).

\bibitem{J-PARC}    
J-PARC proposal P04,
{\it Measurement of high-mass dimuon production at the 50-GeV proton
synchrotron}, J.-C.~Peng and S.~Sawada spokespersons;
{\tt http://j-parc.jp/index-e.html}.

\bibitem{Kumano10}     
S.~Kumano,
J. Phys. Conf. Ser. {\bf 312}, 032005 (2011).

\bibitem{Mul01}
P.~J.~Mulders,      
{\it ``Transverse momentum dependence in structure functions
in hard scattering processes''}, lecture notes,
\texttt{http://www.nikhef.nl/$\sim$pietm/COR-0.pdf}, 2001 (unpublished).

\bibitem{KMPW95}
S.~A.~Kulagin, W.~Melnitchouk, G.~Piller and W.~Weise,
Phys. Rev. C {\bf 52}, 932 (1995).

\bibitem{KMd}
S.~A.~Kulagin and W.~Melnitchouk,
Phys. Rev. C {\bf 78}, 065203 (2008).

\bibitem{Paris}
M.~Lacombe {\it et~al.},
Phys. Lett. B {\bf 101}, 139 (1981).

\bibitem{AV18}
R.~B.~Wiringa, V.~G.~J.~Stoks and R.~Schiavilla,
Phys. Rev. C {\bf 51}, 38 (1995).

\bibitem{CDBonn}
R.~Machleidt,
Phys. Rev. C {\bf 63}, 024001 (2001).

\bibitem{WJC}
F.~Gross and A.~Stadler,
Phys. Rev. C {\bf 78}, 014005 (2008).

\bibitem{FS78}
L.~Frankfurt and M.~Strikman,
Phys. Lett. {\bf 76} B, 333 (1978);
Phys. Rep. {\bf 76}, 215 (1981).

\bibitem{Ethier14}
J.~J.~Ethier, N.~Doshi, S.~Malace and W.~Melnitchouk,
Phys. Rev. C {\bf 89}, 065203 (2014).

\bibitem{EFP83}
R.~K.~Ellis, W.~Furmanski and R.~Petronzio,
Nucl. Phys. {\bf B212}, 29 (1983).

\bibitem{Qiu90}
J.~W.~Qiu,
Phys. Rev. {\bf D 42}, 30 (1990).

\bibitem{AQ08}
A.~Accardi and J.~W.~Qiu,
JHEP {\bf 0807}, 090 (2008).

\bibitem{MSS97}
W.~Melnitchouk, M.~Sargsian and M.~I.~Strikman,       
Z. Phys. A {\bf 359}, 99 (1997).

\bibitem{CJRR}
F.~E.~Close, R.~L.~Jaffe, R.~G.~Roberts and G.~G.~Ross,
Phys. Rev. D {\bf 31}, 1004 (1985).

\bibitem{Ethier13} 
J.~J.~Ethier and W.~Melnitchouk,
Phys. Rev. C {\bf 88}, 054001 (2013).

\bibitem{GRV98} 
M.~Gl\"uck, E.~Reya and A.~Vogt,
Eur. Phys. J. C {\bf 5}, 461 (1998).

\bibitem{JR09} 
P.~Jimenez-Delgado and E.~Reya,
Phys. Rev. D {\bf 79}, 074023 (2009).

\bibitem{JR14} 
P.~Jimenez-Delgado and E.~Reya,
Phys. Rev. D {\bf 89}, 074049 (2014).

\bibitem{Alde90} 
D.~M.~Alde {\it et al.},
Phys. Rev. Lett. {\bf 64}, 2479 (1990).

\bibitem{McGaughey92} 
P.~L.~McGaughey {\it et al.},
Phys. Rev. Lett. {\bf 69}, 1726 (1992).

\bibitem{CTEQ5}
H.~L.~Lai {\it et al.},
Eur. Phys. J. C {\bf 12}, 375 (2000).

\bibitem{Collins88}
J.~C.~Collins, D.~E.~Soper and G.~F.~Sterman,
Adv. Ser. Direct. High Energy Phys. {\bf 5}, 1 (1988).

\bibitem{Peng-priv}
J.-C.~Peng and D.~F.~Geesaman,
private communication.

\bibitem{60fest}
W.~Melnitchouk,
AIP Conf. Proc. {\bf 1261}, 85 (2010).

\bibitem{CJ14}
J.~F.~Owens {\it et al.},
in preparation.

\end{thebibliography}
\end{document}